\documentclass   [12pt,psfig]{article}
\usepackage{amssymb}
\usepackage{amsmath}
\usepackage{a4wide}
\usepackage{epsfig}
\def\t{\tilde}
\def\be{\begin{equation}}
\def\ee{\end{equation}}
\def\bea{\begin{eqnarray}}
\def\eea{\end{eqnarray}}

\def\f#1#2{\frac{#1}{#2}}

\begin{document}
\begin{titlepage}

\title{ \bf On the cosmological backreaction for large distance modifications of gravity} \vskip2in

\author{
{\bf Karel Van Acoleyen$$\footnote{\baselineskip=16pt E-mail: {\tt
karel.vanacoleyen@ugent.be}}}
\hspace{3cm}\\
 $$ {\small Department of Physics and Astronomy, Ghent
 University,}\\
 {\small Krijgslaan 281, S9, 9000 Gent, Belgium}.
}

\date{}
\maketitle
\def\baselinestretch{1.15}
\begin{abstract}
\noindent

Every theory that modifies gravity at cosmological distances and
that is not already ruled out by the Solar system observations
must exhibit some nonlinear mechanism that turns off the
modification close to a compact matter source. Given this
nonlinearity, one might expect such a theory to show a large
gravitational {\em backreaction}, i.e. an order one influence of
the small scale inhomogeneities on the large scale evolution of
the Universe. We argue that this is not necessarily the case. If
the dominant nonlinear terms in the equations obey a shift
symmetry, the averaged effect of the nonlinearities can be small,
although the effect on small scales is large. This happens for DGP(-like) modifications and so called $f(G)$ (or Gauss-Bonnet) models. For both type
of models the shift symmetry is part of the larger "Galilean"
symmetry.

\end{abstract}

\vskip-19cm  \vskip3in

\end{titlepage}
\setcounter{footnote}{0} \setcounter{page}{1}
\newpage
\baselineskip=20pt

\section{Introduction}

The late time acceleration of the Universe can be interpreted in
two qualitatively different fashions, that are both stupefying in
their own right. Either we take it as a manifestation of some
omnipresent dark energy component - dynamical or not- or we view
it as an indication of the breakdown of Einstein gravity at
cosmological distances. It is the latter approach that we want to
discuss in this paper.

Over the years, in fact already since the early days of general
relativity, people have been proposing several possible
modifications of the gravitational Einstein-Hilbert action,
ranging from the inclusion of other curvature invariants or extra
fields with non-minimal couplings to the introduction of extra
dimensions for gravity. The corresponding (4D) modified Einstein
equations can be written generically as: \be
G_{\mu\nu}(g)+H_{\mu\nu}(g,\pi)=8\pi G_N
T_{\mu\nu}\,,\label{modeinstein}\ee  where the modification
appears in the extra tensor $H_{\mu\nu}$ that is added to the
Einstein tensor at the left hand side of the equation. Notice that
we will work in the Jordan-frame and not consider any modifications to the minimal matter-metric
coupling in the matter action that gives rise to the right-hand
side. From the uniqueness of spin-2 gravity \cite{Deser:1987uk} it
follows that the modification will necessarily involve one or more
new degrees of freedom. So if we want to keep the equations at
most of second order in the derivatives, new fields $\pi$ will
appear in addition to the metric field $g_{\mu\nu}$. Actually, in
this paper we will restrict ourselves to those cases and regimes where the
dominant contributions to the modification arise through only one
extra scalar degree of freedom, with one corresponding scalar
field $\pi$. Variation of the action with respect to this new
field then results in one more equation that we have to consider:
\be E_{\pi}(g,\pi)=0\,. \label{pieq}\ee

As a first step in establishing the viability of a particular
model as an alternative for dark energy one then looks at the
evolution of the scale factor in the homogenous (presently matter
dominated) Friedmann-Lema\^{\i}tre-Roberston-Walker (FLRW)
universe to see if there exist self-accelerating solutions. For
those models that produce reasonable fits to the supernovae data,
the transition from an ordinary matter dominated FLRW expansion to
the modified evolution will be controlled by an extra
dimensionfull constant $\mu$\footnote{We work in natural units, $c=\hbar=k_b=1$. }, that is of the order of the present
Hubble parameter $H_0$. A possible parameterization for the modification on the FLRW background then reads: \be H_{00}^{(0)}\sim {H_i^i}^{(0)}\sim \f{\mu^{2n+2}}{H^{2n}}\,,\ee
with $n>1/2$ and the superscript (0) referring to the evaluation on the cosmic background. On the present background we have: $ H_{00}^{(0)}\sim {H_i^i}^{(0)}\sim G_{00}^{(0)}\sim {G_i^i}^{(0)}\sim H_0^2$, and we will normalize the $\pi$-equation (\ref{pieq}) so that also $E_\pi^{(0)}\sim H_0^2$.  Moreover, the modified cosmic evolution implies that on the present cosmic background the extra degree of
freedom is light, with a mass $m \lesssim H_0$, giving it an
effectively infinite interaction range.

Of course, the acceleration of the scale factor in the homogeneous
FLRW background by itself, does not permit us to distinguish
between large distance modifications of gravity and dark energy.
It is only when we consider the inhomogeneous Universe that we
actually live in, that the distinction makes sense. By definition,
for large distance modifications of gravity, the extra light
degree of freedom will couple to matter with gravitational
strength. More explicitly, one can look at the first order terms
of the full equations in an expansion on the cosmic background.
Working in Newtonian gauge and dropping the vector and tensor
perturbations that are subdominant for the scalar matter density
fluctuations in our Universe, we can write the metric for the
Universe as (for simplicity we will consider a flat background):
\be ds^2\approx -dt^2(1+2\psi)+a(t)^2(1+2\phi)(dx^2+dy^2+dz^2)\,.
\label{metric}\ee The rotational symmetry of the FLRW background
then dictates the following general form for the scalar components
of Eqs. (\ref{modeinstein}) and (\ref{pieq}) at linear order: \bea
G^{(1)}_{00}+H^{(1)}_{00}&\approx&-2\f{\nabla^2}{a^2}\phi+
\f{\nabla^2}{a^2}\sigma_1\approx 8\pi G_N\overline\rho{\small(t)}\delta(x,t)\,,\nonumber\\
{G_i^i}^{(1)}\!+{H_i^i}^{(1)}&\approx&2\f{\nabla^2}{a^2}(\phi+\psi)+
\f{\nabla^2}{a^2}\sigma_2\approx0\,, \label{lineq}\\
E_{\pi}^{(1)}&\approx&\f{\nabla^2}{a^2}\sigma_3=0 \nonumber
\,,\eea where the $\sigma_i$ are linear combinations of
$\phi,\psi$ and $\pi$. We will normalize $\pi$ such that $\pi\sim\phi\sim\psi$ on the present cosmic background. The perturbations are sourced by the matter
fluctuations with relative density contrast
$\delta=\Delta\rho/\overline\rho$ ($\overline\rho$ is the average
matter density). We focus on the non-relativistic, subhorizon
density fluctuations, for which the (effective) pressure terms can
be dropped. Accordingly, we have dropped the subdominant terms on
the left hand side. For instance, we do not write possible mass
terms, since as we said, for those models that modify the cosmic
expansion the mass on the present background will be of order
$H_0$. This excludes the so called chameleon models from our
general treatment. For these models the mass of the extra scalar
field is at least of the order $10^3 H_0$, which ties to the fact
that the late time acceleration is simply driven by the
cosmological constant term in the scalar field potential
\cite{Brax:2004qh}. Our treatment does however obviously include
Einstein gravity, for which the $\sigma_i$ are zero. In that case
the first equation is just the Poisson equation for Newton's
potential $\phi$ in comoving coordinates, while the second
equation results in $\psi\approx-\phi$. The parameterization
(\ref{lineq}) also includes DGP (-like), $f(R)$ and $f(G)$ (or Gauss-Bonnet) gravity
\cite{Dvali:2000hr,Carroll:2003wy,Navarro:2005da} as we will show more explicitly in section \ref{specific}. For all those modified gravity models the $\pi$ fluctuation mixes up with the metric fluctuations in the $\sigma_i$,  resulting in an effective gravitational
coupling to matter on the present cosmic background. This direct coupling of the extra degree of freedom with the matter sector,  will necessarily produce a
markedly different phenomenology, once we bring in the
cosmological probes of inhomogeneity (CMB anisotropies, large
scale structure) \cite{Linder:2005in}.

Moreover, the gravitational force is not only important on
cosmological scales, but also on smaller scales. It plays a
prominent role in the dynamics of galaxies, stellar and planetary
systems and as such there are many additional observations that
can be used to constrain possible modifications of gravity . In
this context, the tightest constraints come from those systems
localized on stellar system scales or smaller, where we can safely
neglect any dark components in the energy-momentum tensor. And
according to the observations of binary pulsars, together with the
Solar system observations and the table top experiments on Earth,
there is actually very little room for any modification of gravity
at these distance scales. For the PPN parameter $\gamma\approx
-\phi/\psi$ for instance, a careful observation of the radio waves
emitted between the Cassini satellite and Earth, revealed that for
the gravitational field due to the Sun
$\gamma=1+\mathcal{O}(10^{-5})$, in agreement with the $\gamma=1$
result of ordinary gravity \cite{Bertotti:2003rm}. In contrast,
for the modified gravity models that we discussed above, the extra
scalar polarization in (\ref{lineq}) on the present cosmic
background, yields an order one modification for $\gamma$, in
blatant conflict with this constraint. So these models are ruled
out if the linearization (\ref{lineq}) holds in the Solar system.
Conversely, for any viable large distance modification of gravity,
the Solar system should be in the gravitational non-perturbative
regime \cite{Dvali:2006su}.

The only know mechanism so far, for cosmological large distance (Gpc) modifications of gravity, that decouples the extra degree of freedom close to a compact matter source (like the Sun) is the nonlinear Vainshtein mechanism \cite{Vainshtein:1972sx}. Through this mechanism the linear solution of (\ref{lineq}) breaks down at a characteristic Vainshtein radius $R_V$ from the compact source. This has been well studied for the DGP model \cite{Deffayet:2001uk}. But as we argue in section \ref{ant}, one can anticipate a similar nonlinear effect, with the same Vainshtein radius, for {\em any} cosmological large distance modification of gravity that is not ruled out by the Solar system experiments.

Now, at first sight this nonlinearity seems to come with a high price. As we further argue in section \ref{ant}, any, even mildly nonlinear structure in the Universe will be in the gravitational non-perturbative regime for these models. This seems to pose a serious problem for the cosmological analysis, since one would expect a breakdown of the conventional FLRW framework. Indeed, the working assumption of this framework is that the gravitational effect of all structure can be treated perturbatively \cite{peebles}, which is manifestly not the case anymore.  However, in section \ref{specific}, we show that the precise form of the (gravitational) nonlinearities in the equations, that are large locally, can be such that the averaged effect on large scales, the so called backreaction, is small. This will be the case if the dominant nonlinearities obey a shift symmetry.

As we briefly review in subsection \ref{SECDGP}, this symmetry is known to exist already for the DGP and DGP-like Galileon models. There the symmetry is part of the larger Galilean symmetry \cite{Nicolis:2008in}, that is required for the absence of ghosts. Gauss-Bonnet gravity is also free of ghosts (on certain backgrounds), so one would expect a similar symmetry. This is precisely what we show explicitly in subsection \ref{SECGB}. But before delving into the nonlinear corrections of modified gravity, let us first briefly review the local and global effect of the nonlinear corrections for ordinary gravity.

\section{Nonlinear effects in ordinary gravity}

For Einstein gravity the dominant higher order corrections in the
quasi-static and subhorizon regime will always contain two spatial
derivatives. So at second order for instance, one has the terms
(taking $\psi\sim\phi$): \be G_{00}^{(2)},\,{G^i_i}^{(2)}\sim
\f{1}{a^2}\nabla_i\phi\nabla_i\phi\,,\,\,\phi\f{\nabla^2}{a^2}\phi\,.\label{nonlineinstein}\ee
The local effect of these terms is all too well known. Comparing
them with the first order terms of $G_{00}$ and $G^{i}_i$ in
(\ref{lineq}), one finds that the linear solutions will break down
when $\phi\gtrsim 1$, i.e. in the immediate neighborhood of
neutron stars and black holes. So for Einstein gravity, the
Newtonianly perturbed FLRW metric (\ref{metric}), with the fields
solution of the linearized equations (\ref{lineq}), is nearly
everywhere a good approximation to the actual metric of the
Universe, also for those systems with highly nonlinear density
contrasts \cite{Ishibashi:2005sj}. Galaxy clusters and individual
galaxies for instance have $\delta \sim 1-10$, $\delta\sim 10^5$
respectively, while the potential typically has the value
$\phi\sim 10^{-5}$. To be clear, the dynamics of these nonlinear systems with respect to the FLRW background is very much non-perturbative, but from the gravitational point of view they are still in the perturbative regime, well described by the linearized equations (\ref{lineq}).

As for the global effect of the terms (\ref{nonlineinstein}), one
should compare their spatial average with the zero order terms
$G_{00}^{(0)},{G^{i}_i}^{(0)}\!\sim H^2$ that appear in the
Friedmann equations. This average can be evaluated from the
knowledge of the density power spectrum. For the modes with
wavelength $L$, we have $\phi_L\sim L^2 H^2\delta$ from the
Poisson equation, so that their contribution can be estimated as:
\be <\phi\nabla^2\phi> \sim H^4L^2|\delta_L|^2 \,\,(\sim 4\pi
G_N<\!\phi_L\Delta\rho_L\!>)\,,\ee with $|\delta_L|^2$ the
dimensionless spectral density. The last term here illustrates its
physical meaning as the contribution to the Friedmann equations of
the average gravitational potential energy of the inhomogeneities.
And for our present Universe this contribution is of relative
order $ 10^{-5}$ which justifies the FLRW framework for cosmology,
that treats the Universe as approximately homogeneous at all
scales \cite{Siegel:2005xu}.

\section{Anticipating nonlinear effects in modified gravity}\label{ant}

Before looking at the particular models just yet, let us now give
a heuristic argument, complementing other general arguments
\cite{Nicolis:2008in,Lue:2003ky}, on the type of higher order
terms we would generically expect for cosmologically large
distance modifications of gravity. Different from Einstein                                                                                                                                           gravity, the modified equations will be nonlinear functions of
curvature. So, in addition to terms of the type
(\ref{nonlineinstein}) that arise in the weak field expansion of
curvature, we should also expect terms that are due to the
expansion of these nonlinear functions. In particular, on cosmic
backgrounds with curvature larger or of the same order as the
characteristic scale of the model, $H^2\gtrsim \mu^2$, this
expansion will break down for local curvatures that are larger
than the background curvature, since this is the only relevant scale in the expansion. Estimating the local curvature due
to inhomogeneity as $H^2\delta$, this means that the weak field
expansion should break down for all systems with nonlinear density
contrasts. This is indeed what happens for DGP gravity \cite{Lue:2004rj} and Galileon gravity \cite{Nicolis:2008in}. As we show in subsection \ref{SECGB} this also happens for $f(G)$ gravity. It does not happen for $f(R)$ gravity,
the reason being that for those models the relevant curvature
($R$) due to an inhomogeneity is locked to its background value
\cite{Erickcek:2006vf}. So for $f(R)$ models the linearization
(\ref{lineq}) holds in the same situations as for ordinary
gravity, and the contribution of the extra light scalar in the
Solar system rules these models out \footnote{
This statement refers to those $f(R)$ models that modify gravity at
cosmological distance scales. Viable $f(R)$ models are equivalent
to the chameleon models that modify gravity (only) up to distances of 1
Mpc or smaller \cite{Faulkner:2006ub}.}.

As we said, for DGP(-like) and Gauss-Bonnet gravity, the range of validity of
the linearization will be different than for Einstein gravity.
From the argument above, we can write down the general form of the
extra nonlinear terms on the present background as \footnote{We limit ourselves to the case where the breakdown of linearity is induced by the second order terms in the equations. The generalization to higher order terms is straightforward.  } (taking $\phi\sim\psi\sim\pi$ and not
being precise about the exact derivative structure): \be
H_{00}^{(2)},\,{H^i_i}^{(2)},E_\pi^{(2)}\sim
\f{\nabla^2\phi\nabla^2\phi}{a^4 H_0^2}\,.\label{nonlinmod}\ee
Indeed, comparing these terms with the first order terms
(\ref{lineq}), we see that the linearization breaks down for
$\nabla^2\phi\sim H_0^2\delta\gtrsim H_0^2$.

The local effect of these terms for an isolated compact object of
mass $M$ can be characterized by the so called Vainshtein radius
\cite{Vainshtein:1972sx}. Within a 'halo' of radius $\sim R_V$
that is surrounding the object, the linear solution for the
gravitational field will break down. We can estimate this radius
easily, again by comparing the individual linear terms in
(\ref{lineq}) with the second order terms (\ref{nonlinmod}), now
for the linear solution $\phi\sim G_N M/(ar)$. With the individual
first order terms of the form:   \be
G_{00}^{(1)},\,{G^i_i}^{(1)}\,,H_{00}^{(1)},\,{H^i_i}^{(1)},E_\pi^{(1)}\sim
\f{G_N M}{(ar)^3}\,, \ee and the dominant second order terms of
the form: \be H_{00}^{(2)},\,{H^i_i}^{(2)},E_\pi^{(2)}\sim \f{(G_N
M)^2}{(ar)^6 H_0^2}\,, \ee we find that the latter become of the
order of the former for \be ar \sim R_V \equiv (G_N
M/H_0^2)^{1/3}\,.\label{Vainshtein}\ee This agrees with the explicit
calculation for the DGP model \cite{Deffayet:2001uk,Lue:2004rj} and the Galileon-models \cite{Nicolis:2008in}. And from our general argument it should also agree with the
situation for any cosmological large distance modification of
gravity that is not ruled out by the Solar system experiments. As
we show explicitly in subsection \ref{SECGB}, the dominant
nonlinear terms for Gauss-Bonnet gravity are indeed of the general
form (\ref{nonlinmod}), causing a breakdown of the linear solution
at the same $R_V$ (\ref{Vainshtein}).

Of
course, having a breakdown of the linear solution at $r\leqslant R_V$ by itself is not enough to ensure the screening of
the extra scalar and therefore the recovery of ordinary gravity at short distances. However, for DGP gravity,
full nonlinear solutions have been obtained that match the cosmic
background at infinity and for which Einstein gravity is indeed recovered
at short distances from the object \cite{Lue:2004rj} . For Gauss-Bonnet
gravity one can also argue the recovery of Einstein gravity at
short distances, although the precise matching to the cosmic
background at infinity is still an open issue
\cite{Navarro:2005gh}.

The global effect of the new nonlinear terms in our inhomogeneous
Universe has been largely left unexplored so far, and in fact for
all studies of the cosmology for these models it was (often
implicitly) assumed that, just as for Einstein gravity, the small
scale structure plays a negligible role. (See \cite{DGPcosmology},\cite{Chow:2009fm,Kobayashi:2009wr,Galileoncosmology} and \cite{Gauss-Bonnetcosmology} for a list of cosmological studies of the DGP model, the DGP-like Galileonmodel(s) and the Gauss-Bonnet models, respectively.) However, for these models, the present Universe consists of compact objects
and structures separated by distances that are smaller or of the
same order as their Vainshtein radius. So in stark constrast with the situation for ordinary gravity, the linearization
(\ref{lineq}) will in fact \emph{never hold locally} and one would not a
priori expect our Universe to expand in the same fashion as the
corresponding homogeneous matter dominated FLRW universe. As a
first step in addressing this issue, we will only consider those
density fluctuations which are still in the linear regime, or on
the brink of going nonlinear $\delta\lesssim 1$. The advantage
being that we can still use the linear solution (\ref{lineq}) that
relates the fields to the densities. In the same way as we did for
ordinary gravity, we can then estimate the contribution of the
modes with wavelength $L$ to the average of the nonlinear terms
as:  \be \f{<\nabla^2\phi\nabla^2\phi>}{H_0^2}\sim
H_0^2|\delta_L|^2\,. \label{antcosmo} \ee  Comparing this average, with the zero order terms that are all three of order $H_0^2$ with our conventions, we find that already mildly nonlinear
density fluctuations $\delta_L\rightarrow 1$ seem to induce an
order one backreaction. This corroborates our expectations since
for those density fluctuations the corresponding 'gravitational
energy density' in (\ref{nonlinmod}) becomes of the same order as
their bare density. However, so far we have not been explicit on
the precise form of the nonlinear terms. In fact, it turns out
that both for DGP(-like) and Gauss-Bonnet gravity the contributions to the
averaged equations of terms of the form (\ref{nonlinmod}) cancel
out; or that the full average gravitational energy is small, of
the same order as for Einstein gravity. As we will show
explicitly in the next section, the reason behind this lies in the shift symmetry for
the relevant terms in these models.

\section{Specific examples}\label{specific}
\subsection{DGP and Galileon gravity}\label{SECDGP}
It has been shown that in the subhorizon regime the 5D DGP model  is effectively described by a 4D theory with
one extra scalar $\pi$ that corresponds to the bending mode \cite{Nicolis:2004qq}, which
inspired the formulation of fully 4D DGP-like models, the so called Galileon models
\cite{Nicolis:2008in}. For the DGP-model, the $\sigma_i$ in the linearized equations
(\ref{lineq}) read: \be
2\sigma_1=-\sigma_2=\pi\,,\,\,\,\,\sigma_3=-4\phi-2\psi\pm
6\f{H}{\mu}(1+\f{\dot{H}}{3H^2})\pi\,,\label{DGPlin}\ee with the minus sign for the
self-accelerating solution \cite{Koyama:2005kd}.\footnote{We have actually rederived these equations from the results of \cite{Nicolis:2004qq}. Our $\pi$ is their $m \pi$.  In \cite{Koyama:2005kd}, the three linearized scalar DGP-equations are presented in a different format, but one can check that our expressions (\ref{DGPlin}) are equivalent.} So the linearization is indeed of the form (\ref{lineq}), with an order one coupling of $\pi$ to the matter density perturbations on the present cosmic background ($\mu\sim H_0$). Similar expressions hold for the full scalar-tensor Galileon models \cite{Chow:2009fm,Kobayashi:2009wr}.

For DGP(-like) gravity the higher order terms have been widely discussed
in the literature (see \cite{Deffayet:2001uk,Nicolis:2004qq,Nicolis:2008in} and citations
therein). They are generated by the term in the Lagrangian
(dropping the overall $M_p^2$ factor and with our normalization
for $\pi$): \be \mathcal{L}_\pi=-\f{1}{\mu^2}\Box \pi
\partial_{\mu}\pi\partial^{\mu}\pi\,.\label{galileon}\ee The corresponding second order correction to the equation $E_\pi$ then indeed has the scaling that we anticipated in (\ref{nonlinmod}). But there is quite some symmetry. Besides having the obvious shift symmetry on $\pi$,
(\ref{galileon}) is also invariant (upon partial integration and dropping
subdominant derivatives of $a$) under a constant shift of the
derivatives of $\pi$: \be \pi \rightarrow \pi+c+c_\mu x^\mu\,. \ee
This so called "Galilean" invariance is a consistency requirement,
since it prevents the appearance of a new degree of freedom that
has to be a ghost by Ostrogradski's theorem \cite{Woodard:2006nt}.
Indeed, the symmetry dictates that, although the Lagrangian
contains four derivatives, the equation that follows from it will
only contain second order derivatives, and no new degrees of
freedom appear \footnote{
But notice that on certain backgrounds the $\pi$ excitation itself
can still be ghostlike \cite{Nicolis:2008in}. A similar comment
applies to $f(G)$ gravity \cite{DeFelice:2006pg}.}
.

However, the normal shift symmetry by itself is sufficient to
ensure a cancellation of the different averaged terms. Since by
this symmetry the terms in the equation $E_\pi$, that follow from
the Lagrangian (\ref{galileon}), appear as a divergence of the
associated Noether current. Explicitly, in the quasi-static and
subhorizon regime, the relevant terms are: \bea
E_\pi^{(2)}&=&\f{2}{\mu^2a^4}\partial_i\left(\nabla^2\pi\partial_i\pi-\partial_i\partial_j\pi\partial_j\pi\right)\nonumber\\
&=&
\f{2}{a^4\mu^2}\left(\nabla^2\pi\nabla^2\pi-\partial_i\partial_j\pi\partial_i\partial_j\pi\right)\,.\label{Epidgp}\eea
In the setup (\ref{metric}), where the infinite wavelength modes
are put in the background, the average of a (spatial) divergence
is zero. As one can easily verify, for finite wavelength modes,
the contributions to the average of the two terms on the last line
of (\ref{Epidgp}) cancel out. For the Galileon-models \cite{Nicolis:2008in}, one also considers a possible fourth and fifth order Galileo-invariant term in the Lagrangian, besides the third order term (\ref{galileon}). These will of course also vanish upon averaging.

\subsection{Gauss-Bonnet gravity}\label{SECGB}
The Gauss-Bonnet models are defined by adding some
function $f(G)$ of the Gauss-Bonnet invariant, \be G\equiv
R^2-4R_{\mu\nu}R^{\mu\nu}+R_{\mu\nu\rho\sigma}R^{\mu\nu\rho\sigma}\,,
\ee to the Einstein-Hilbert term $R$ of the gravitational
Lagrangian. The choice for this invariant is precisely motivated
by the fact that the corresponding higher derivative equations
only show one extra degree of freedom in addition to the spin-2
graviton \cite{Navarro:2005da}. This can be seen easily by
eliminating the higher order derivatives through the introduction
of a Lagrange multiplier $\lambda$ \cite{DeFelice:2006pg}: \be
\mathcal{L}=\f{M_{p}^2}{2}\big(R+f'(\lambda)(G-\lambda)+f(\lambda)\big)\label{GB}
\,.\ee In this way it is evident that the models are in fact
equivalent to a scalar field with some potential (third and fourth
term), no canonical kinetic term and a non-minimal coupling to $G$
(second term). Due to the specific total derivative structure of
$G$, this non-minimal coupling will generate at most second order
derivatives (both of $\lambda$ and the metric fields) in the full
equations \cite{DeFelice:2006pg}. Notice that $\lambda=G$ by the $\lambda$-equation.

We can read off the $\sigma_i$ in (\ref{lineq}) for the  FLRW linearization of Gauss-Bonnet gravity from the full linearized equations of \cite{DeFelice:2009ak}\footnote{Explicitly, from the Eqs. 36-38 in that paper one can read off $H_{00}^{(1)}$ and ${H_{i}^i}^{(1)}$, putting $\alpha=\psi$, $\beta=\gamma=0$, $F=1$, $H^2\delta\xi=\pi$\, and using $\partial_j {H^{ji}}^{(1)}=0$ in the quasi-static subhorizon limit. The $E_\pi^{(1)}$ equation follows from their Eq. 46.}:

\be \sigma_1=-8H\dot{\xi}\phi-4\pi,\,\,\sigma_2=-8H\dot{\xi}\psi-8\ddot{\xi}\phi-8(1+\f{\dot{H}}{H^2})\pi
,\,\,\sigma_3=(1+\f{\dot{H}}{H^2})\phi+\psi\,.\ee
 We use here $\xi\equiv f'(\lambda_0)$, where $\lambda_0$ is the FLRW background value of $\lambda$ (or $G$). The dots denote time-derivatives. Our scalar degree of freedom now represents the fluctuations of the Gauss-Bonnet invariant, $\pi\equiv
H^2f''(\lambda_0)\t{\lambda}$, with  $\tilde{\lambda}=\lambda-\lambda_0$. With the general estimate of  $f(\lambda_0)$ and its derivatives on the present cosmic background as: \be f^{(n)}(\lambda_0)\sim H_0^{2-4n}\,,\ee one can show that the mass terms in the full equations of \cite{DeFelice:2009ak} are subdominant, with an effective mass of the order $H_0$ on the present background. So,  we find indeed a linearization of the form (\ref{lineq}), with an order one effective gravitational coupling for the extra scalar. Notice that the dominant subhorizon terms in $H_{00}^{(1)},\,{H^i_i}^{(1)}$ and $E_\pi^{(1)}$, which are the terms with the maximum number of spatial derivatives, all arise from the non-minimal coupling term in (\ref{GB}), since it is the only term that contains derivatives.

The precise form of the higher order terms for Gauss-Bonnet gravity has
so far not been discussed in the literature. We will focus here on
those terms that are relevant to the central theme of the paper,
leaving a more detailed discussion for future work. For the same reason as before, it will be the non-minimal coupling term that is responsible for the dominant higher order corrections. In the
quasi-static subhorizon limit, the dominant contribution to the third order expansion of this term reads (again dropping $M_p^2$): \be
\mathcal{L}^{(3)}_{\lambda
G}=\left(f'(\lambda)G\right)^{(3)}\approx f''(\lambda_0)\t{\lambda}G^{(2)}=\f{8}{a^4 H^2}\pi\left(
\nabla^2\phi\nabla^2\psi-\partial_i\partial_j\phi\partial_i\partial_j\psi\right)\,.\label{GBnonlin}\ee And again, as in the case of DGP and Galileon gravity, we find the anticipated scaling behavior (\ref{nonlinmod}) of the corresponding second order equations. But also now we expect this term to have the full Galilean symmetry,
since the equations that derive from it only contain second
derivatives. And indeed, up to boundary terms: \bea \f{a^4
H^2}{8}\mathcal{L}^{(3)}_{\lambda G}&=&\phi\left(
\nabla^2\psi\nabla^2\pi-\partial_i\partial_j\psi\partial_i\partial_j\pi\right)\nonumber\\
&=&\psi\left(\nabla^2\pi\nabla^2\phi-\partial_i\partial_j\pi\partial_i\partial_j\phi\right),
\label{GBnonlin2}\eea which shows explicitly that the Lagrangian is invariant under
a shift of all three fields and their derivatives separately. The
shift symmetry by itself again guarantees that the terms are divergences that vanish upon averaging. One can read off these terms explicitly,
now appearing in all three equations $H_{00}$, $H^i_i$ and $E_\pi$, from the second and first line of the right-hand side of (\ref{GBnonlin2}) and the right-hand side of (\ref{GBnonlin}), respectively.

\section{Conclusions}
To conclude, let us summarize the main results of the paper. First
of all we have argued on general grounds, that for those models
that modify gravity at cosmological (Hubble) distance scales and
that are not ruled by the Solar system tests, the linearization on the present
FLRW background (\ref{lineq}) will break down for nonlinear
density contrasts, $\delta\geq 1$. For an isolated compact object
of mass $M$ this translates to a universal Vainshtein radius
$R_V\sim (G_N M/H_0^2)^{1/3}$ at which the linearization breaks
down. Our argument is heuristic, but it complements other
arguments \cite{Nicolis:2008in,Lue:2003ky}, and more importantly it is confirmed by the
specific examples. For the DGP and Galileon models the breakdown
of the linearization is already well studied. This is not the case
for the Gauss-Bonnet models, for which we have explicitly written
down the dominant nonlinear term in the action for the first time (\ref{GBnonlin}).

As such, the status of the FLRW linearization (\ref{lineq}) is
very different than in the case of Einstein gravity, since now the
linearization will in fact never hold locally for our present
Universe. Correspondingly one might expect a large cosmological
backreaction for these models, as confirmed by our (naive) scaling
argument (\ref{antcosmo}). But as we have shown further on, the scaling argument does not capture possible cancellations when the nonlinearities are averaged. In that case the conventional FLRW framework would still be viable. This is precisely what happens when there is a shift symmetry. This symmetry is part of the larger Galilean symmetry that is required for consistency of the models under consideration. And as we have shown explicitly, the Galilean symmetry, which was originally recognized for the DGP model, is indeed also present for the dominant terms in the Gauss-Bonnet models.

We should stress that we have only considered mildly nonlinear universes with density contrasts $\delta \lesssim 1$. Showing that the FLRW framework remains valid in a more realistic universe, with highly nonlinear density fluctuations
$\delta\gg 1$, requires going beyond our perturbative approach. In
this light it is interesting to note that the very same shift
symmetry is used to argue the equivalence principle for
 \mbox{DGP(-like)} gravity \cite{Hui:2009kc}. We can view the validity of the conventional FLRW framework as a cosmological version of the equivalence principle.

\section{Acknowledgments} It's a pleasure to thank Eanna Flanagan, Ignacio Navarro and Aseem Paranjape for their comments on a first version of the paper. I am supported by the Fund for Scientific
Research-Flanders (Belgium).

\end{document}